\documentclass[%
 aip,
amsmath,amssymb,
reprint,%
]{revtex4-1}

\usepackage{graphicx}
\usepackage{dcolumn}
\usepackage{bm}

\usepackage[utf8]{inputenc}
\usepackage[T1]{fontenc}
\usepackage{mathptmx}
\usepackage{etoolbox}

\usepackage{lineno}
\usepackage[separate-uncertainty = true,multi-part-units=single,per-mode=symbol]{siunitx}  
\usepackage{color}
\usepackage{pdfcomment}
\usepackage{graphicx}
\usepackage{mwe}
\usepackage{tikz}
\usepackage[caption=false]{subfig}

\makeatletter
\def\@email#1#2{%
 \endgroup
 \patchcmd{\titleblock@produce}
  {\frontmatter@RRAPformat}
  {\frontmatter@RRAPformat{\produce@RRAP{*#1\href{mailto:#2}{#2}}}\frontmatter@RRAPformat}
  {}{}
}%
\makeatother
\begin{document}

\preprint{AIP/123-QED}

\title[Cryogenic Optical-to-Microwave Conversion Using Si Photonic Integrated Circuit Ge Photodiodes]{Cryogenic Optical-to-Microwave Conversion Using Si Photonic Integrated Circuit Ge Photodiodes}
\author{D. Julien-Neitzert}
\affiliation{Department of Electrical and Computer Engineering, The University of British Columbia, Vancouver, Canada}
\affiliation{Quantum Matter Institute, The University of British Columbia, Vancouver, Canada}
\author{E. Leung}
\affiliation{Department of Electrical and Computer Engineering, The University of British Columbia, Vancouver, Canada}
\affiliation{Quantum Matter Institute, The University of British Columbia, Vancouver, Canada}
\author{N. Islam}
\affiliation{Department of Electrical and Computer Engineering, The University of British Columbia, Vancouver, Canada}
\affiliation{Quantum Matter Institute, The University of British Columbia, Vancouver, Canada}
\author{S. Khorev}
\affiliation{Department of Electrical and Computer Engineering, The University of British Columbia, Vancouver, Canada}
\affiliation{Quantum Matter Institute, The University of British Columbia, Vancouver, Canada}
\author{S. Shekhar}
\affiliation{Department of Electrical and Computer Engineering, The University of British Columbia, Vancouver, Canada}
\author{L. Chrostowski}
\affiliation{Department of Electrical and Computer Engineering, The University of British Columbia, Vancouver, Canada}
\affiliation{Quantum Matter Institute, The University of British Columbia, Vancouver, Canada}
\author{Jeff F. Young}
\affiliation{Quantum Matter Institute, The University of British Columbia, Vancouver, Canada}
\affiliation{Department of Physics and Astronomy, The University of British Columbia, Vancouver, Canada}
\author{J. Salfi}
\email{jsalfi@ece.ubc.ca}
\affiliation{Department of Electrical and Computer Engineering, The University of British Columbia, Vancouver, Canada}
\affiliation{Quantum Matter Institute, The University of British Columbia, Vancouver, Canada}
\affiliation{Department of Physics and Astronomy, The University of British Columbia, Vancouver, Canada}

\date{\today}

\begin{abstract}
Integrated circuit technology enables the scaling of circuit complexity and functionality while maintaining manufacturability and reliability. Integration is expected to play an important role in quantum information technologies, including in the highly demanding task of producing the classical signals to control and measure quantum circuits at scales needed for fault-tolerant quantum computation. Here we experimentally characterize the cryogenic performance of a miniaturized photonic integrated circuit fabricated by a commercial foundry that down-converts classical optical signals to microwave signals. The circuit consists of waveguide-integrated germanium PIN photodiodes packaged using a scalable photonic wire bonding approach to a multi-channel optical fiber array that provides the optical excitation. We find the peak optical-to-microwave conversion response to be $\sim 150 \pm 13$~mA/W in the O-band at 4.2~K, well below the temperature the circuit was designed for and tested at in the past, for two different diode designs. The second diode design operates to over 6~GHz of 3~dB bandwidth making it suitable for controlling quantum circuits, with improvements in bandwidth and response expected from improved packaging. The demonstrated miniaturization and integration offers new perspectives for wavelength-division multiplexed control of microwave quantum circuits and scalable processors using light delivered by optical fiber arrays.
\end{abstract}

\maketitle

\section{\label{introduction}Introduction}

Quantum computers, in which information is encoded and processed in collections of two-level quantum systems called qubits, have the potential to perform calculations that are intractable on classical computers in real-world problems\cite{Motta2021, Cao2019,Schuld2014,Rebentrost2018}. It is well-known that carrying out a classically intractable calculation on a quantum computer requires high precision manipulation of up to several hundred thousand to a million physical qubits\cite{Cleland2012,von2021quantum}. The large number of physical qubits is associated with reliable manipulation of quantum information employing fault-tolerant circuits and quantum error correction\cite{Cleland2012,von2021quantum}. 

The combination of precise qubit operations and large qubit count presents serious difficulties in the implementation of quantum hardware. One difficulty is that many quantum hardware platforms, including superconducting qubits\cite{Krantz2019}, spin qubits\cite{Burkard2023}, and readout circuits for photonic qubits\cite{Bussieres2014}, operate at temperatures between approximately 4.2~K and 10~mK. In each case, the low temperatures reduce the impact of noise, and for superconducting qubits and photonic qubit readout circuits, enables operation below the critical temperature of the superconducting materials used. One challenge for operating more than a few hundred operated qubits is the heat load associated with the electrical connection between ambient control electronics and qubits\cite{krinner2019engineering}. Typically, quantum circuits operating in the microwave regime employ one semi-rigid coaxial line per qubit to bring control pulses from room temperature electronics into a cryostat. The thermal conductance of the semi-rigid coaxial lines is such that their thermal load alone places a limit of approximately 1,500 coaxial lines at a 4~K cryocooler stage\cite{Lecocq2021}, well below the number of qubits likely to be needed for useful fault-tolerant quantum computation. 
 
Potential solutions to the challenge of controlling large numbers of qubits include synthesizing the control circuits in-situ using cryogenic silicon-based logic\cite{bardin2019design} or superconducting flux-based logic\cite{takeuchi2010chip, mcdermott2018quantum}. Another potential solution more recently identified is to employ optical fibers, which have negligible thermal conductance, to carry signals in the optical domain which can be down-converted to the microwave regime at low temperatures\cite{Lecocq2021,Joshi2022,Li2024}. Optical fibers have much larger bandwidths than coaxial cables, such that wavelength division multiplexing could be used to carry many control signals on a single fiber. For this to result in a net advantage, the heat load of the demultiplexing and optical-to-microwave conversion modules inside the cryostat must be less than the all-microwave alternative.  Previous work has shown that discrete PN photodiodes can be used for down-conversion, and could potentially offer low dissipation\cite{Lecocq2021, Li2024, Joshi2022}. Scalable control would benefit from many diodes and wavelength division demultiplexing filters all operating at low temperature on a photonic integrated circuit. However, to date, very little is known about the behaviour of photonic-integrated-circuit diodes at low temperatures. The responsivity, frequency response, and I-V characteristics of discrete\cite{Bardalen2018} and flip-chipped\cite{bardalen2016packaging} InGaAs/InP photodiodes have been measured at 4~K. The I-V characteristics of waveguide integrated diodes have been measured down to 34~K for Ge\cite{Pizzone2020} and low frequency photoresponse has been measured down to 77~K for GeSn\cite{bansal2024temperature}. Unconventional defect-based light-emitting diodes have been integrated on waveguides with single-photon detectors\cite{buckley2017all}. Recent work has demonstrated the use of silicon-photonics based electro-optic modulators at 4K for microwave-to-optical conversion\cite{Yin21}. However, integrated circuit optical-to-microwave conversion at temperatures low enough for operation of superconducting circuits has received virtually no attention.  

Here we experimentally demonstrate the cryogenic operation of miniaturized Ge PIN photodiodes (GePD) on the silicon on insulator (SOI) photonic-integrated-circuit (PIC)  platform including down-conversion of microwave-modulated optical signals to the microwave regime. The PIC is fabricated by a commercial foundry\cite{Novack2013,Fard2016}. Our integrated GePDs\cite{Novack2013,Fard2016} are packaged using polymer photonic wirebonds\cite{lindenmann2012photonic,lindenmann2015connecting} (PWB) connecting discrete fibers in a ribbon array to Si waveguides that currently offers the highest density of optical connections, which is highly favourably for circuit scaling. On the microwave side, the chip was packaged using conventional electrical wirebonds. We experimentally characterize the dark current, optical wavelength-dependent response, and microwave frequency response of the PWB-waveguide-GePD subsystem for optical to microwave downconversion.

\section{\label{device_and_packaging}Device and Packaging}

\begin{figure}[ht!]
  \begin{minipage}[b]{0.38\textwidth}
    \centering
    \begin{tikzpicture}
      \node[inner sep=0pt] at (0,0) {\includegraphics[width=\linewidth]{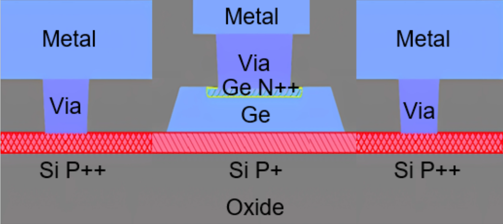}};
      \node[inner sep=0pt] at (-90pt, -37pt) {(a)};
    \end{tikzpicture}
  \end{minipage}
  \hfill
  \begin{minipage}[b]{0.38\textwidth}
    \centering
    \begin{tikzpicture}
      \node[inner sep=0pt] (b) at (0,0) {\includegraphics[width=\linewidth]{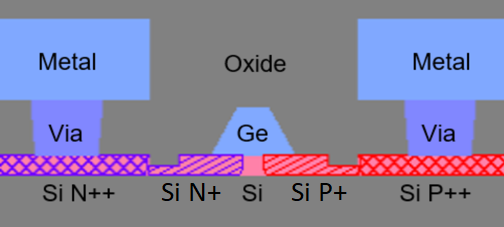}};
      \node[inner sep=0pt] at (-90pt, -37pt) {(b)};
    \end{tikzpicture}
  \end{minipage}
\caption{(a) vertical GePD defined by an shallow N++ doped germanium grown epitaxially on a silicon doped P+ layer. (b) floating GePD defined epitaxially grown germanium on top silicon layer with P+ and N+ layers with proximate  P++ and N++ regions. In the floating GePD, the Ge does not make direct contact to the P++ and N++ silicon. Both GePDs are based on designs described in ref \cite{Fard2016, Novack2013,Zhang2014}.}
\label{cross_sections}
\end{figure}

We investigated two different integrated GePD structures on SOI, both illustrated in Figure \ref{cross_sections}, referred to as vertical\cite{Fard2016, Novack2013} and floating\cite{Zhang2014} GePDs. The vertical GePD has N++ and P+ regions in Ge and Si, respectively, that are arranged vertically. The floating GePD has N+ and P+ regions arranged laterally in silicon, with a Ge layer on top. The lateral dimensions of the vertical and floating GePDs including the highly doped contacts are $14 \times 21$~$\mu$m and $21 \times 15$~$\mu$m, respectively. The devices were fabricated on two separate chips using the Advanced Micro Foundry (AMF) silicon photonic process. More details on the device fabrication are given in Appendix \ref{sec:manufacturing}.    



\begin{figure}[ht!]
  \begin{minipage}[b]{0.38\textwidth}
    \centering
    \begin{tikzpicture}
      \node[inner sep=0pt] at (0,0) {\includegraphics[width=\linewidth]{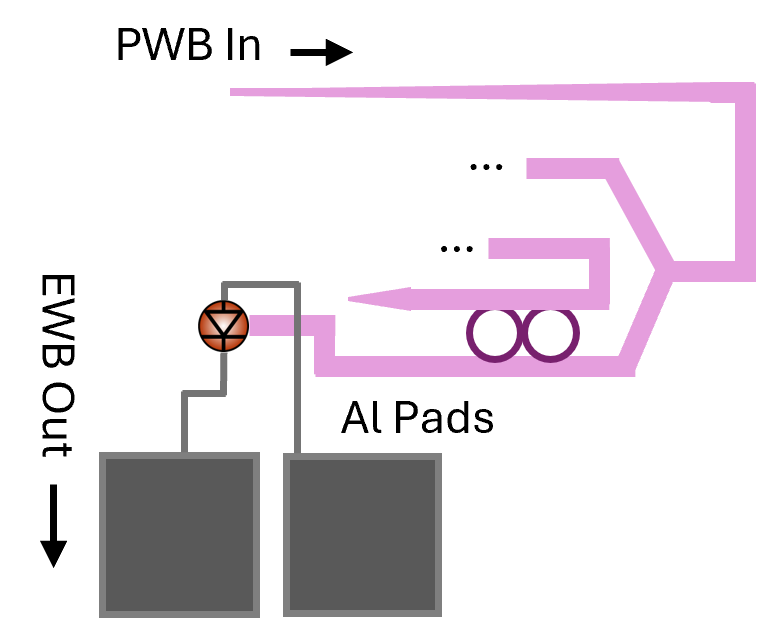}};
      \node[inner sep=0pt] at (70pt, -60pt) {\includegraphics[width=0.25\linewidth]{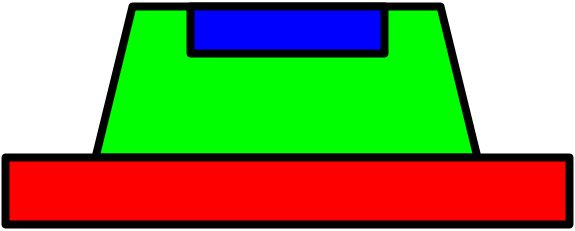}};
      \node[inner sep=0pt] at (70pt, -75pt) {\it{Vertical}};
      \node[inner sep=0pt] at (-80pt, 60pt) {(a)};
    \end{tikzpicture}
  \end{minipage}
  \hfill
  \begin{minipage}[b]{0.38\textwidth}
    \centering
    \begin{tikzpicture}
      \node[inner sep=0pt] (b) at (0,0) {\includegraphics[width=\linewidth]{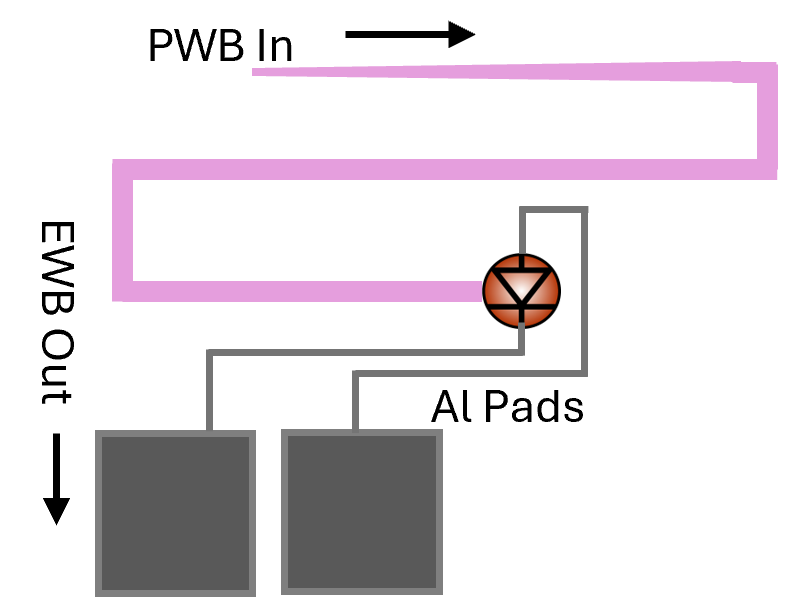}};
      \node[inner sep=0pt] (d) at (70pt,-57pt) {\includegraphics[width=0.25\linewidth]{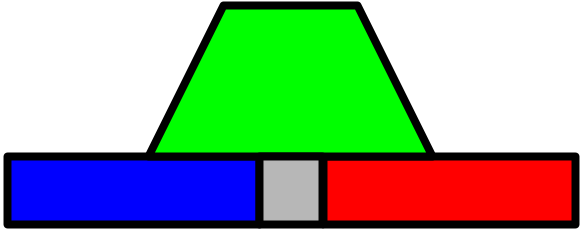}};
      \node[inner sep=0pt] at (70pt, -72pt) {\it{Floating}};
      \node[inner sep=0pt] at (-80pt, 60pt) {(b)};
    \end{tikzpicture}
  \end{minipage}
\caption{A schematic of the SOI chips which contain the GePDs as well as other integrated photonic devices. For both GePD assemblies a photonic wire bond (PWB) couples light from a fiber array to on-chip silicon routing waveguides  employing a 65~$\mu$m long surface coupling nanotaper with a width varying from 140~nm to 500 nm. EWBs connect the pads on chip to the PCB. (a) The vertical GePD is integrated on a waveguide with a multiplexing section that includes a y-branch. The multiplexing section has a termination waveguide on the add port of the rings while the through port leads into our photodiode. (b) The floating GePD chip  has a shorter, direct waveguide connection between the PWB and the diode. 
}
\label{on_chip}
\end{figure}
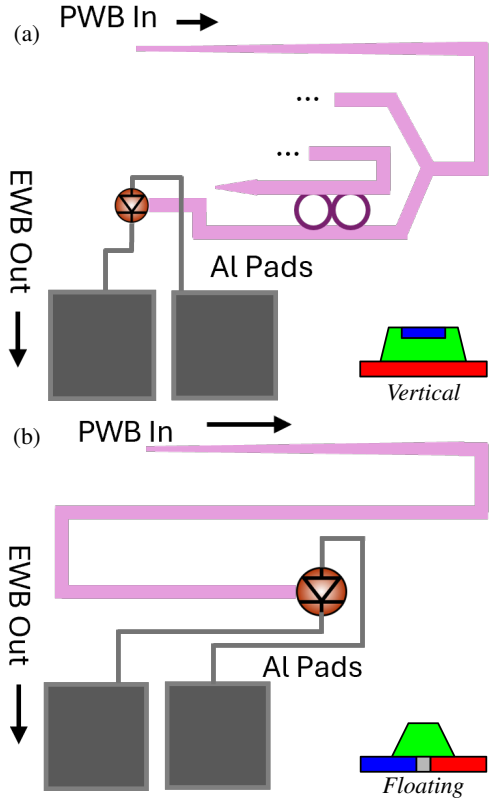

The photonic and electrical connections for the two different GePDs are illustrated in Figure \ref{on_chip}. After dicing the wafer pieces supplied by AMF, the integrated circuits were packaged on nominally identical assemblies with fiber arrays (FAs) and electrical printed circuit boards (PCBs) connected to the chips by PWBs and electrical wirebonds (EWBs), respectively. The optical path in the circuit with the vertical GePD includes microring resonators providing wavelength add/drop functionality that would be useful in future multi-qubit control experiments, and which are absent in the circuit for the floating GePD. The long surface coupling nanotapers have a length of 65~$\mu$m with a tapered width of 140 to 500~nm. Low (2 dB) losses on similar PWBs have recently been demonstrated at 4.2~K \cite{Lin2023}. 

The SOI chip containing the circuits in Fig.~\ref{cross_sections} was mounted using a procedure similar, but not identical to, the procedure described in reference ~\onlinecite{Lin2023}. First, a UV glue was used to fix the SOI chip onto a small area silicon shim, which was then glued onto a PCB with 50-ohm grounded coplanar waveguides. The silicon shim was used to add a height offset to the SOI chip such that the difference in height between the core of the FA and the SOI chip is below 75~$
\mu$m. This helps to reduce the loss that comes from bends in the 3D printed PWB. A micro-positioner was used to laterally align the FA to the surface tapers on chip in order to minimize bending loss and to optimize the PWB's efficiency. After the alignment, the FA was glued onto the silicon shim. Master Bond EP29LPSPAO-1 Black epoxy was then applied to areas of the assembly that do not have the UV glue applied. When the epoxy cured, the UV glue was removed and the PWBs were formed between the FA and the on-chip tapers. The EWBs are 20~$\mu$m diameter Al and connect the bondpads of the GePD to the 50-ohm PCB waveguide. These assemblies are illustrated schematically in Figure \ref{fig:experimental_setup}, and photos are given in Figure~\ref{fig:assembly_irl} in Appendix~\ref{sec:manufacturing}. The main difference compared to the previous study was the use of a Si shim in this work, versus a Cu shim in previous work\cite{Lin2023}. One limitation in the chip design is a lack of a loopback structure for isolating and characterizing PWB loss.

\section{\label{device_characterization}Device Characterization}

\begin{figure}[ht!]
    \setlength{\fboxsep}{3pt} 
    \includegraphics[width=0.5\textwidth]{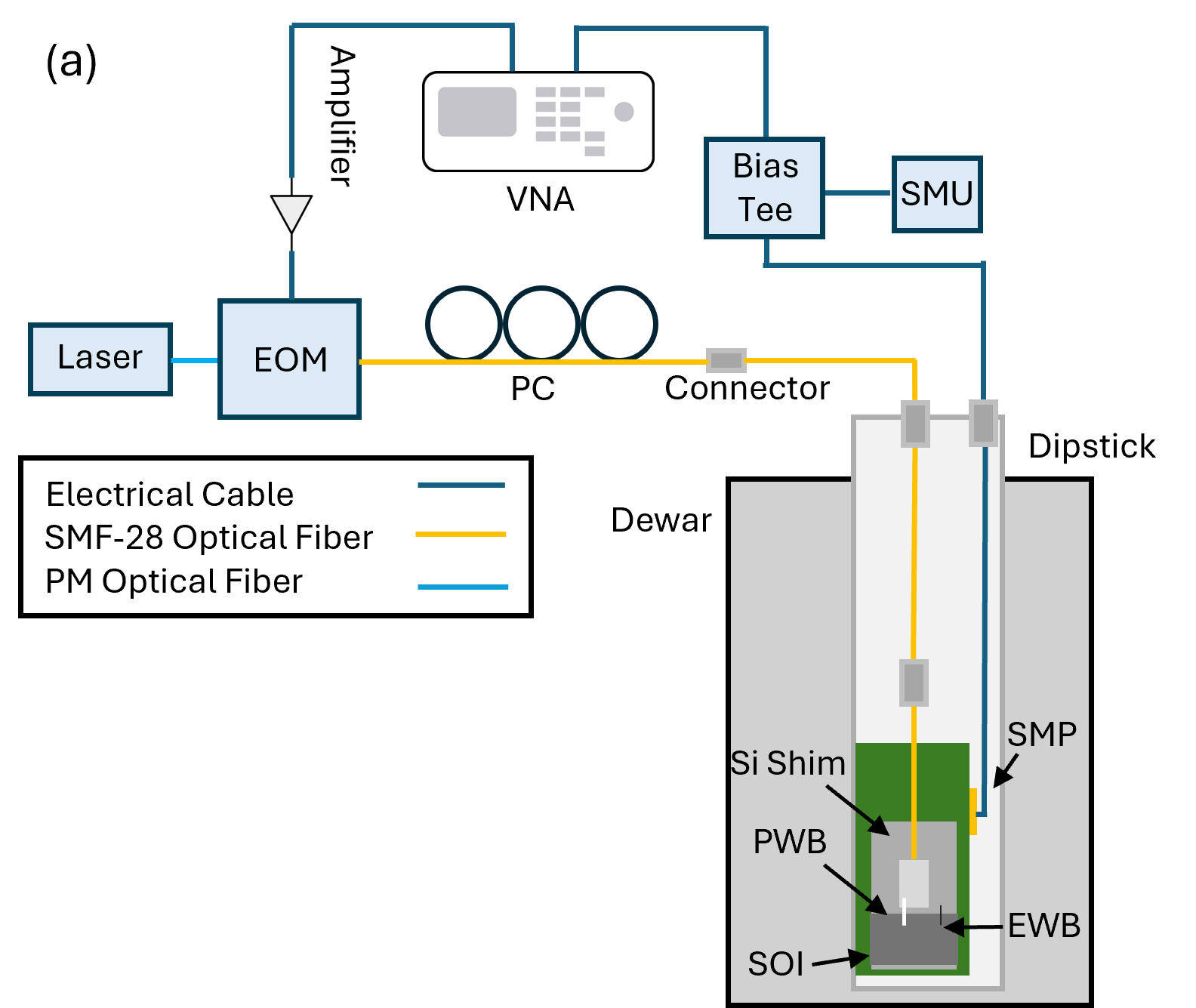}\llap{\raisebox{0.145cm}{\makebox[0.515\textwidth][l]{\fbox{\includegraphics[height=0.13\textwidth]{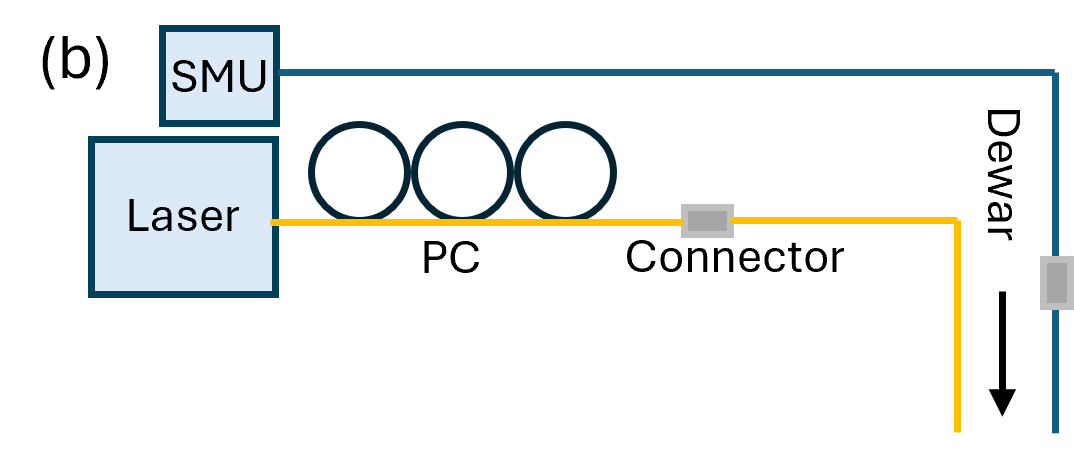}}}}}
\caption{(a) Experimental setup for optical-microwave conversion. An electro-optic modulator (EOM) is used to modulate light from a laser using a continuous microwave tone from port 1 of a vector network analyzer (VNA), after it is amplified to a suitable level. The VNA measures the downconverted signal from the GePD on port 2. A source measure unit (SMU) applies a bias and measure low-frequency current through a bias tee. (b) Experimental setup for I-V measurements and low-frequency photo response. A source measure unit (SMU) applies a bias and measure low-frequency current. The optical output from our laser goes through a polarization controller (PC) to maximize responsivity. In both setups, a single mode fiber is used to carry the optical signal into and out of the polarization controller (PC), and a polarization maintaining fiber (PMF) is employed just before the dipstick fiber connector.}  
\label{fig:experimental_setup}
\end{figure}

Each assembly was measured separately in a dipstick filled with a few mbar of He exchange gas, fitted with both microwave and optical fiber vacuum feedthroughs connected to CuNi 50-ohm coax and a 1550~nm single-mode fiber. Experiments were performed in the dipstick at room temperature and 4.2~K as measured using a calibrated Allen Bradley resistor\cite{schulte1966carbon}. Two different room-temperature experimental setups were used, as shown in Figure \ref{fig:experimental_setup}: one for measuring the I-V characteristics and the low-frequency optical responsivity of the GePDs, and another for optical-to-microwave down-conversion using an optical signal modulated using an electro-optic-modulator (EOM) by microwave tones with frequencies up to 10~GHz. Two different tunable lasers were used for the measurements, one covering the O-band wavelengths from approximately 1250~nm to 1350~nm and another covering the C-band wavelengths from approximately 1450~nm to 1550~nm. Several devices were measured, first at room temperature, then at 4.2~K. The temperature was cycled from room temperature to 4.2~K a few times, to judge the reproducibility of the overall assembly performance, including the PWB and GePD. 

\subsection{\label{iv_characteristics}GePD I-V Characteristics}

\begin{figure}[ht!]
  \begin{minipage}[b]{0.38\textwidth}
    \centering
    \begin{tikzpicture}
      \node[inner sep=0pt] at (0,0) {\includegraphics[width=\linewidth]{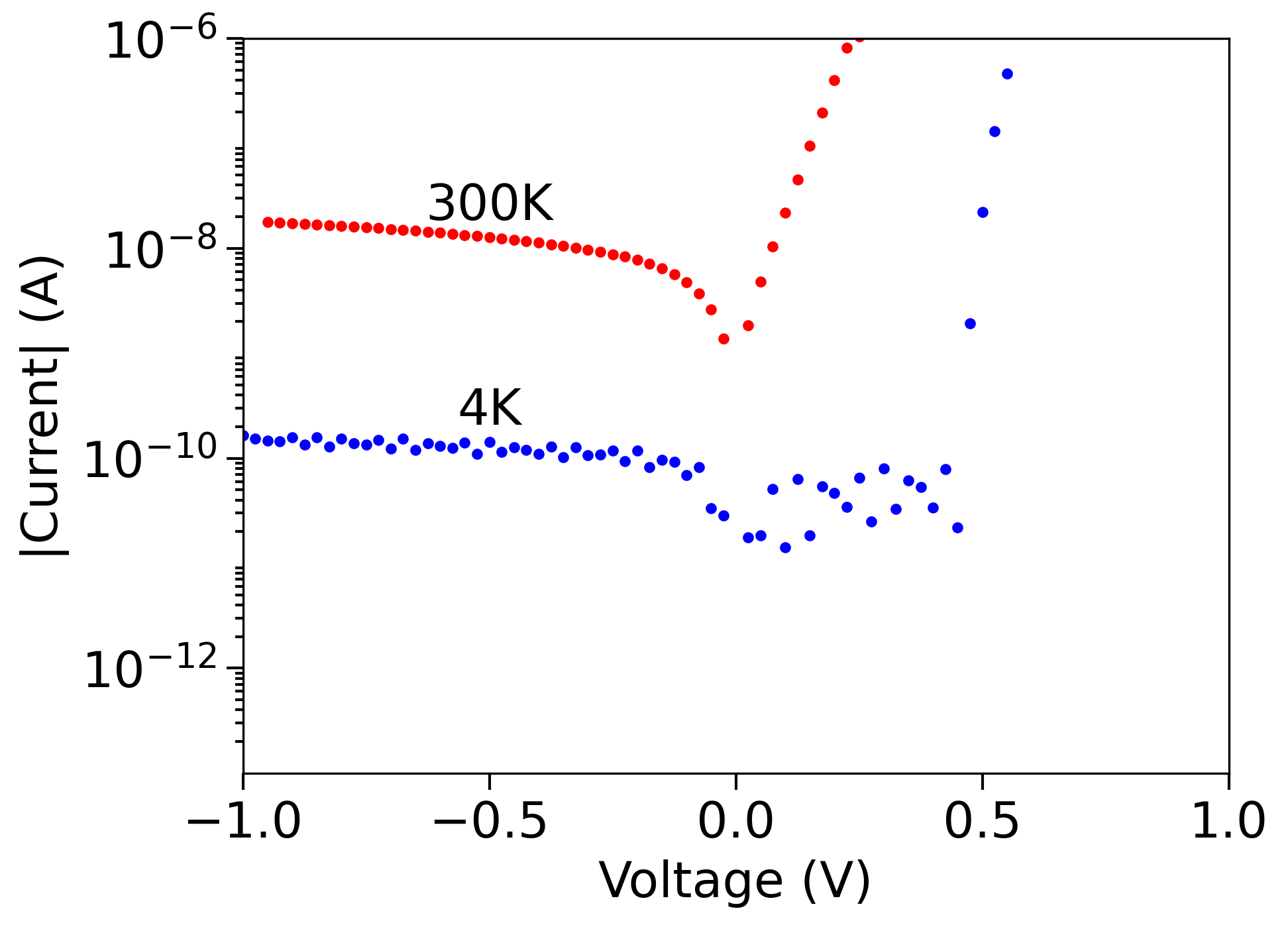}};
      \node[inner sep=0pt] at (-25pt,-25pt) {\includegraphics[width=0.25\linewidth]{Figures/rect_cartoon.png}};
      \node[inner sep=0pt] at (-25pt, -40pt) {\it{Vertical}};
      \node[inner sep=0pt] at (-92pt, 60pt) {\bf{(a)}};
    \end{tikzpicture}
  \end{minipage}
  \hfill
  \begin{minipage}[b]{0.38\textwidth}
    \centering
    \begin{tikzpicture}
      \node[inner sep=0pt] (b) at (0,0) {\includegraphics[width=\linewidth]{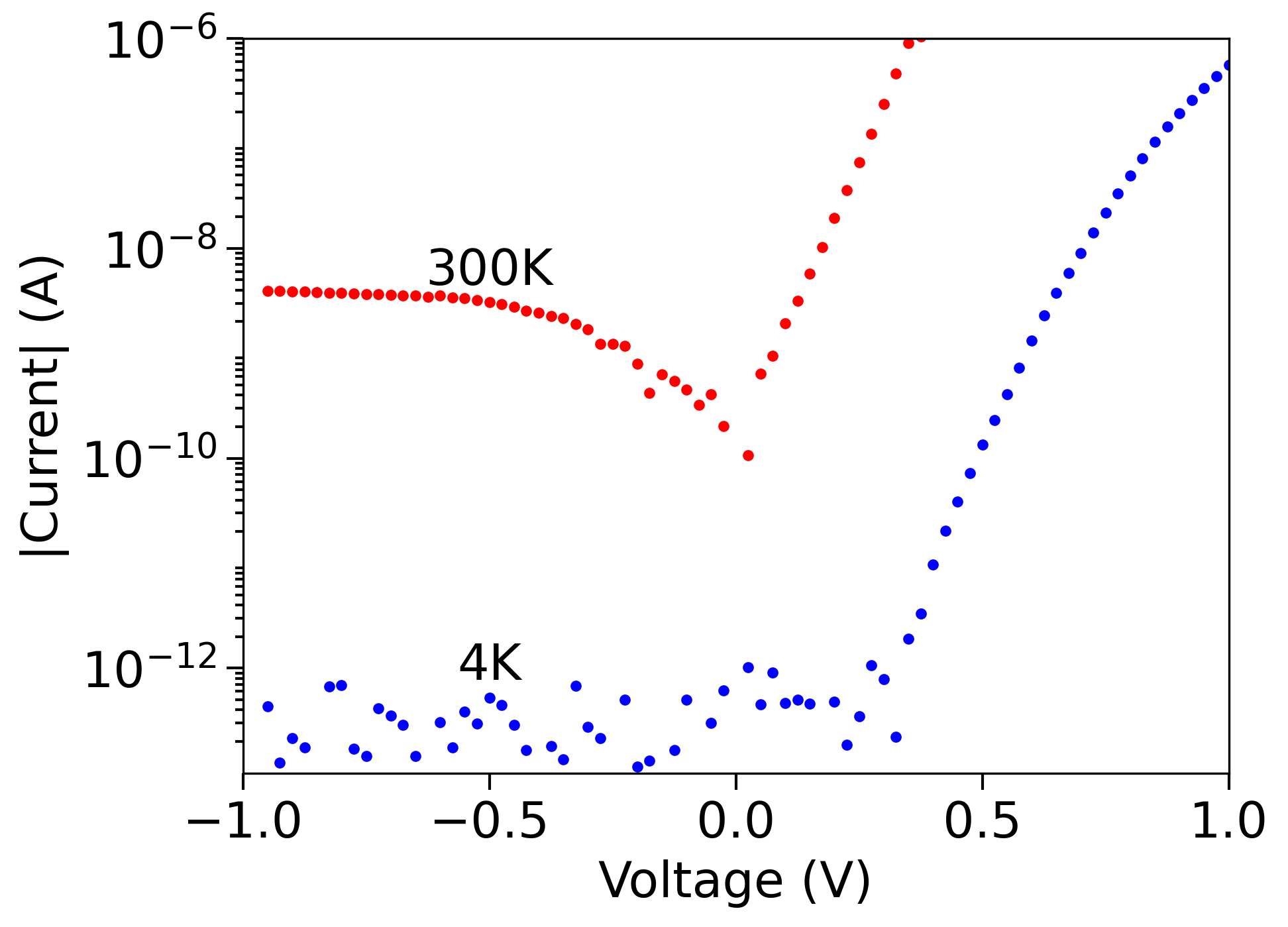}};
      \node[inner sep=0pt] (d) at (-25pt,0pt) {\includegraphics[width=0.25\linewidth]{Figures/float_cartoon.png}};
      \node[inner sep=0pt] at (-25pt, -15pt) {\it{Floating}};
      \node[inner sep=0pt] at (-92pt, 60pt) {\bf{(b)}};
    \end{tikzpicture}
  \end{minipage}
  \caption{Dark I-V characteristic of the PWB-GePD subsystem at 300~K and 4.2~K for (a) the vertical and (b) floating GePD }
  \label{fig:iv}
\end{figure}

The dark I-V characteristics of both the vertical and floating GePDs were measured at room temperature and 4.2~K, using the setup in Figure \ref{fig:experimental_setup}b. Figure \ref{fig:iv} shows the result of two representative GePDs. Both the vertical and floating GePDs have reverse bias dark currents comparable to previous work at room temperature \cite{Zhang2014,Pizzone2020} which decrease to approximately 0.1~nA (1~pA) at 4.2~K for the vertical (floating) GePD. The I-V curve shows that both GePDs maintain their rectifying behaviour at 4.2~K. The rectification of the GePD at 4.2~K indicates that the doping of both GePDs is high enough to avoid freeze out of the charge carriers. As expected, the dark current is much lower for both the vertical and floating GePD at 4.2~K where there is substantially less thermal generation in the depletion region. The reverse dark current may also be impacted by changes to carrier diffusivity and concentrations in the highly doped regions of the GePD at low temperature, which temperature-dependent experiments could elucidate. The floating GePD I-V characteristic is noticeably sub-exponential in forward bias, which is expected for a diode with significant series resistance. In comparison, the vertical GePD turns on more abruptly and has a more exponential characteristic, as expected at low temperature if the series resistance is negligible. Factors that might impact the series resistance include the hetero-interfaces between the Ge and Si regions, and the different device shapes and doping profiles (Figure~\ref{fig:iv}). One major difference is that the vertical GePD has a N++ germanium doped region, while the floating GePD has a N+ silicon region.

\subsection{\label{optical_response_section}Low-Frequency Optical Response}

\begin{figure}[ht!]
  \begin{minipage}[b]{0.38\textwidth}
    \centering
    \begin{tikzpicture}
      \node[inner sep=0pt] at (0,0) {\includegraphics[width=\linewidth]{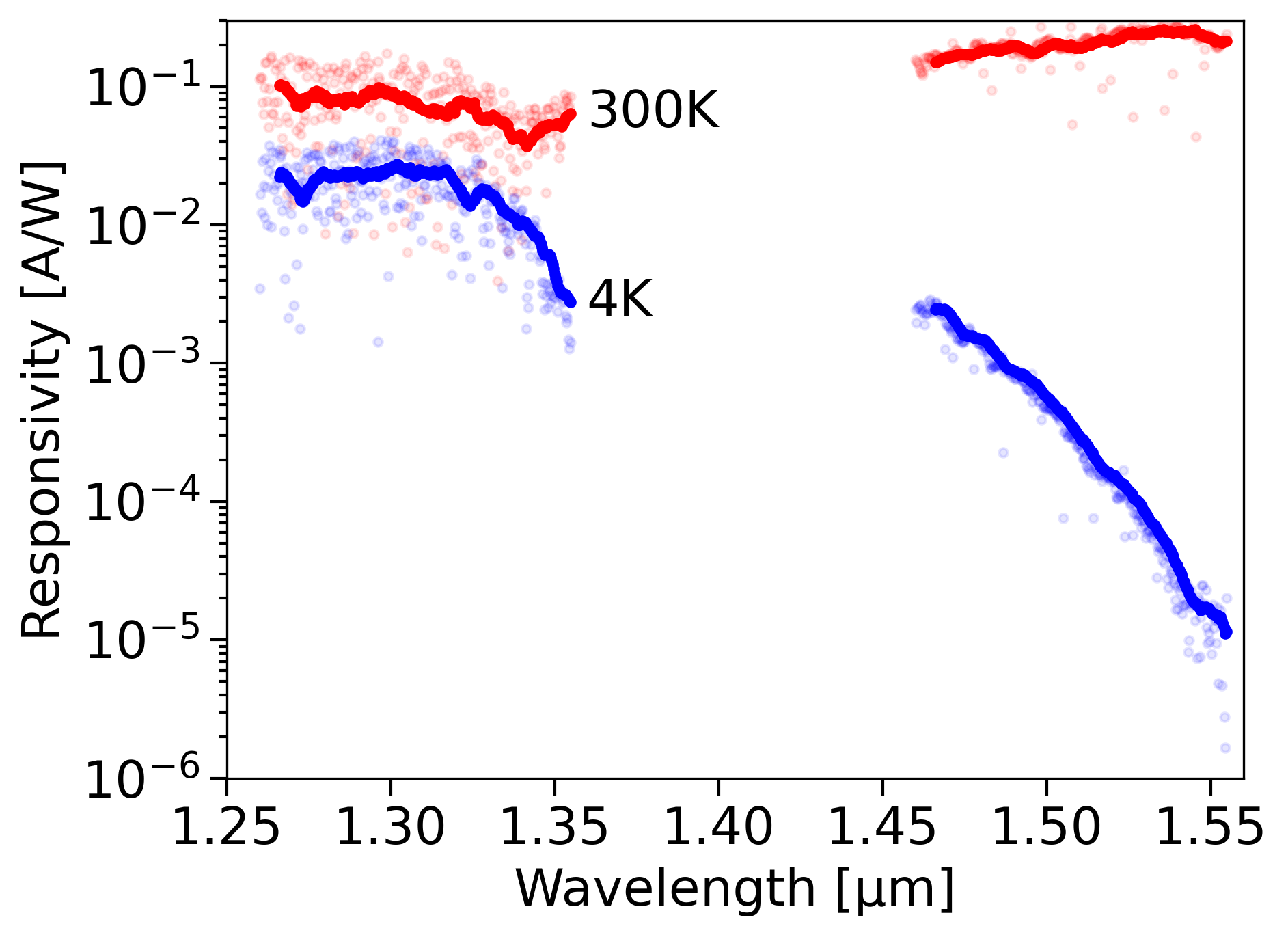}};
      \node[inner sep=0pt] at (-25pt,-25pt) {\includegraphics[width=0.25\linewidth]{Figures/rect_cartoon.png}};
      \node[inner sep=0pt] at (-25pt, -40pt) {\it{Vertical}};
      \node[inner sep=0pt] at (-92pt, 65pt) {\bf{(a)}};
    \end{tikzpicture}
  \end{minipage}
  \hfill
  \begin{minipage}[b]{0.38\textwidth}
    \centering
    \begin{tikzpicture}
      \node[inner sep=0pt] (b) at (0,0) {\includegraphics[width=\linewidth]{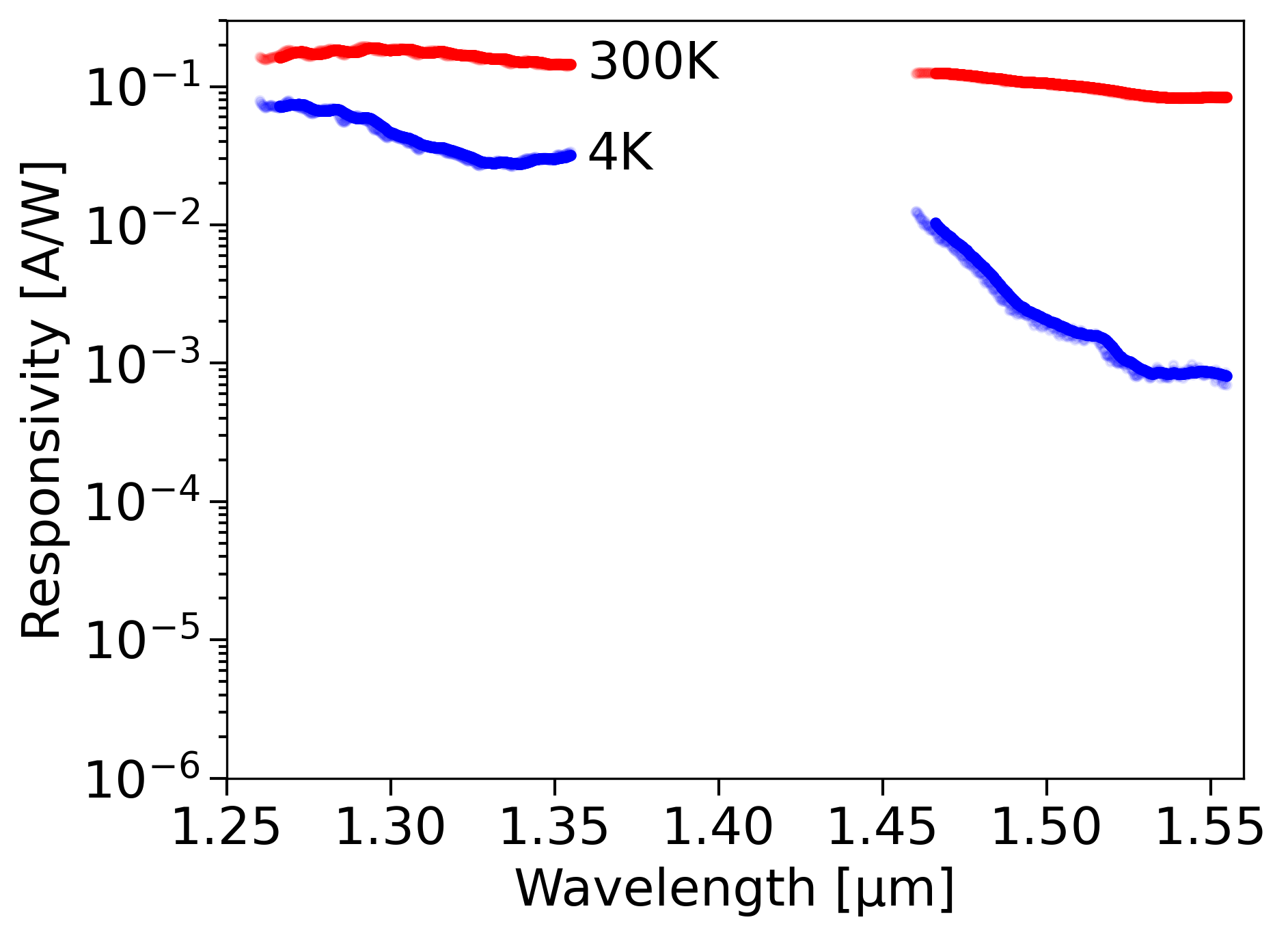}};
      \node[inner sep=0pt] (d) at (-25pt,-25pt) {\includegraphics[width=0.25\linewidth]{Figures/float_cartoon.png}};
      \node[inner sep=0pt] at (-25pt, -40pt) {\it{Floating}};
      \node[inner sep=0pt] at (-92pt, 65pt) {\bf{(b)}};
    \end{tikzpicture}
  \end{minipage}
  \caption{Photocurrent per unit optical excitation power of the PWB-GePD subsystem at 300~K and 4.2~K for (a) vertical GePD and (b) floating GePD. Solid red and blue points: filtered data using a 20-point moving average filter. Light red and blue points: raw data.}
  \label{fig:dc}
\end{figure}

The DC current per unit of optical power was measured using the setup shown in Figure~\ref{fig:experimental_setup}b. Table \ref{tab:loss_optical} summarizes the sources of and estimated values and uncertainties for the optical losses between the reference plane (optical fiber feedthrough) and the GePD, for both GePDs.  The net optical losses, excluding those attributable to the PWB, are $5.78 \pm 0.2$~dB and $1.47 \pm 0.2$~dB ($6.08 \pm 0.2$~dB and $1.47 \pm 0.2$~dB) in the O-band (C-band) for the vertical and floating GePD chips respectively. The measured photocurrent divided by the incident optical power, corrected for the estimated optical path losses, results in the net estimated spectral responsivity of the GePD assembly including PWB losses reported in Figure \ref{fig:dc}. Inspecting the data for both GePDs, we see that the responsivity of the vertical GePD fluctuates with wavelength much more than the floating GePD, which has a smoother responsivity.  This is expected, for two reasons. First, the vertical GePD chip has microring resonators coupled to the waveguide channel connecting the PWB to the GePD, which introduced numerous, deep and narrowband dropouts in the transmitted laser power that are evident in both O and C band data.  Second, in the O-band the waveguides are not single mode, so light coupled from the PWB transits to the diode via more than a single mode. These modes have slightly different phase propagation constants and transverse mode profiles, so the net field distribution where the waveguide meets the Ge detector varies as a function of wavelength due to mode beating, thus affecting the coupling efficiency to the diode.  This introduces a more gradual but still significant variation of the absorbed laser power as a function of the optical wavelength in Fig.~\ref{fig:dc}a. This effect is reduced in the floating chip data because the connecting waveguide is much shorter. As reported in Figure~\ref{fig:dc}, the O-band response is significantly higher than the C-band response at 4.2~K. This is expected, based on the temperature dependence of the direct band gap of germanium, that varies from $0.805 \pm 0.001$$~eV to $$0.898 \pm 0.001$~eV from 300~K  to 4.2~K  \cite{Zwerdling1959}.  This translates to a wavelength shift from 1540 ~nm to 1380~nm. The reverse bias had a negligible impact on the response in the bias range of-0.5~V to 0~V. 

The peak net responsivities are $250\pm 11$~mA/W and $270\pm 12$~mA/W for the floating and vertical GePDs, respectively, at room temperature. These values are somewhat smaller than values of approximately 660 mA/Watt to 1090 mA/Watt reported in the literature\cite{Fard2016}. The peak values of responsivity at 4.2 ~K, $80 \pm 4$~mA/W and $40\pm 2$~mA/W respectively for the floating and vertical GePDs are somewhat lower than the values measured at room temperature. 

Several factors may be responsible for the suppression of the responsivity.  Optical loss in the PWB could change with temperature due to differential thermal contraction of elements in the assembly, that may increase bending losses. The previous work on cryogenic PWBs by reference \onlinecite{Lin2023} with a reported loss of 2~dB uses a Cu shim to reduce bending losses that arise from thermal contraction. Additional losses from thermal contraction mismatch between our assembly is possible and is likely a cause of additional optical loss in our system. The intrinsic responsivity of the  GePD could decrease at lower temperatures due to, \textit{e.g.}, changes to the GePD depletion region width, strain, or excess carrier properties at low temperature. We note that the data reported in Figure~\ref{fig:dc} was obtained first at room temperature and during the first cooldown. The room temperature responsivity after the first temperature cycle reduced by around a factor of three (See Appendix for details), and remained fixed on subsequent cooldowns, suggesting that some of the decrease observed at low temperature may be due to a permanent change in either the PWB or possibly the GePD upon its first cooldown to 4.2 ~K. Further study is required to investigate causes for the change in device performance with temperature cycling, and to understand possible temperature dependent behaviour of the PWB and GePD. Uncertainty in the response due to mode beating, that we remove by averaging, is difficult to quantify, and not included in the reported error bars. Most importantly, this data shows that Si-integrated GePDs remain functional photodetectors at temperatures of $4.2$~K. 

The wavelength dependence of the responsivity is slightly different for the vertical and floating GePDs. Despite similar peak responsivities of $80 \pm 4$~mA/W  and $40\pm 2$ mA/W, the roll-off of the responsivity in the low-temperature data towards longer wavelengths is more severe for the vertical GePD. At 1550~nm, where the GePD weakly absorbs at 4.2~K due to the Ge band-gap shift, the responsivity is only 10 $\mu$A/W for the vertical GePD versus 1~mA/W for the floating GePD.  

\subsection{\label{optical_to_microwave}Optical-to-Microwave Downconversion}
The frequency response of optical-to-microwave downconversion was measured using the setup illustrated in Figure \ref{fig:experimental_setup}a, in the O-band 1308~nm), away from microring resonator dropout, and where appreciable optical absorption occurs in Ge at 4.2~K. For both assemblies, the frequency response of lossy components and amplifying components in the off-chip microwave path was characterized and subtracted from the measured data. This loss calibration includes, on the output microwave line, frequency dependent losses for the CuNi semi-rigid cables, the room temperature SMA cables, and the bias tee (Figure~\ref{fig:experimental_setup}a). On the circuit feeding the EOM, we take into account frequency dependent losses for the room temperature SMA cables, and the conversion of microwave signals to modulated light by the EOM. Microwave losses are summarized in Table~\ref{tab:loss_microwave}. Calibration the optical losses follows the same approach as for the low-frequency measurements in Sec.~\ref{optical_response_section}. The excitation is measured at a reference plane, which is the fiber feedthrough of the cryostat, losses are estimated for remaining optical components up to, but not including, the PWB. The modulation efficiency of light at the output of the EOM had to be determined, which was done by measuring the change in phase as a function of its bias voltage. By performing this calibration of the EOM, the photoresponse for the optical-microwave downconversion can be considered as a composite of the PWB, GePD and electrical wire bonds, and can be directly compared to low-frequency photoresponse. Laser power levels between 100~$\mu$W to 1~mW were employed. To properly compare measurements the laser power in each experiment was normalized by subtracting a factor of $10 \log_{10}((\qty{20}{\micro\watt}/P_{\text{optical}}[\unit{\micro\watt}])^2)$ resulting in a frequency response in dB per 20 ~$\mu$W of laser power. 

\begin{figure}[ht!]
  \begin{minipage}[b]{0.38\textwidth}
    \centering
    \begin{tikzpicture}
      \node[inner sep=0pt] at (0,0) {\includegraphics[width=\linewidth]{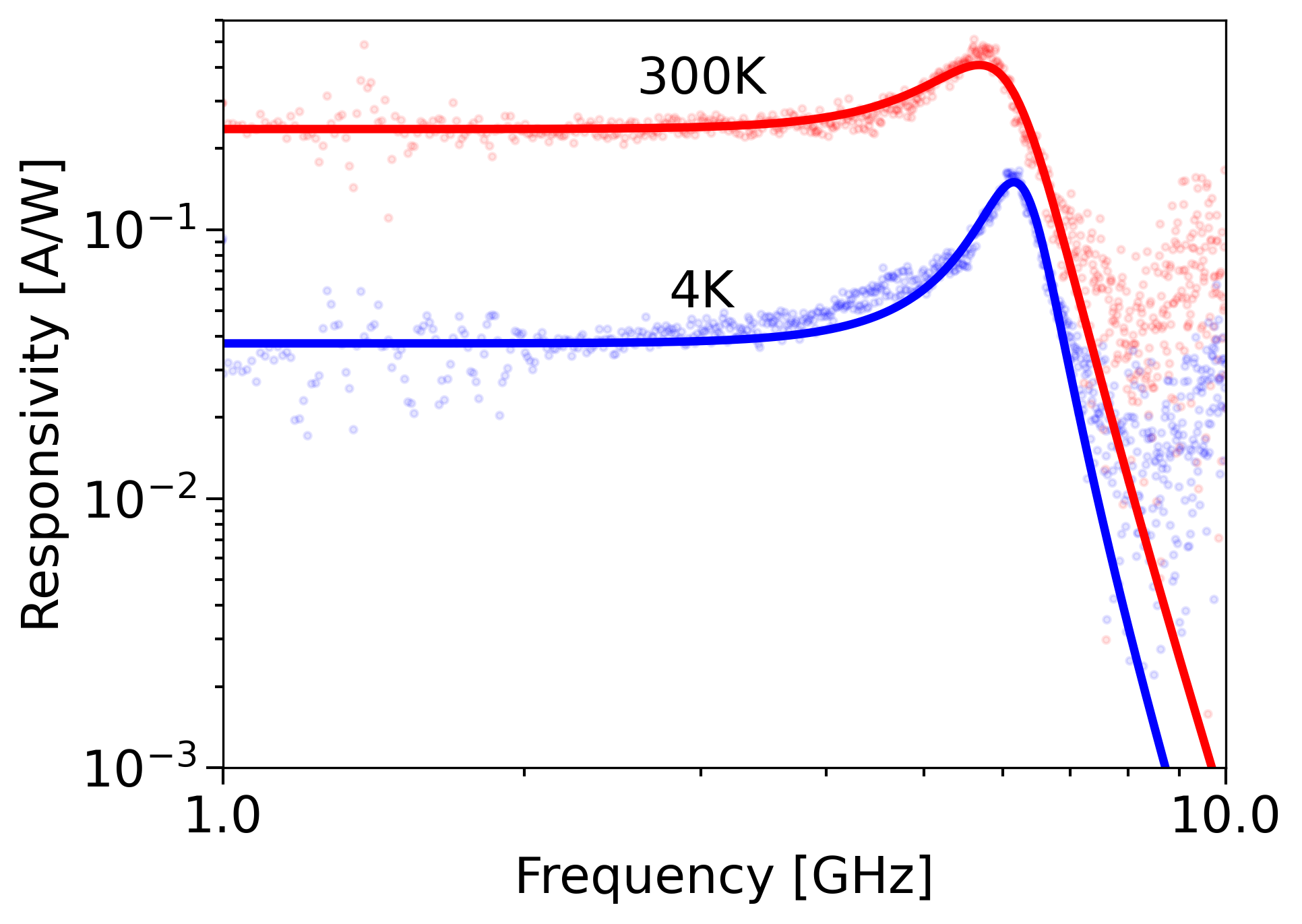}};
      \node[inner sep=0pt] at (-25pt,-25pt) {\includegraphics[width=0.25\linewidth]{Figures/rect_cartoon.png}};
      \node[inner sep=0pt] at (-25pt, -40pt) {\it{Vertical}};
      \node[inner sep=0pt] at (-92pt, 60pt) {\bf{(a)}};
    \end{tikzpicture}
  \end{minipage}
  \hfill
  \begin{minipage}[b]{0.38\textwidth}
    \centering
    \begin{tikzpicture}
      \node[inner sep=0pt] (b) at (0,0) {\includegraphics[width=\linewidth]{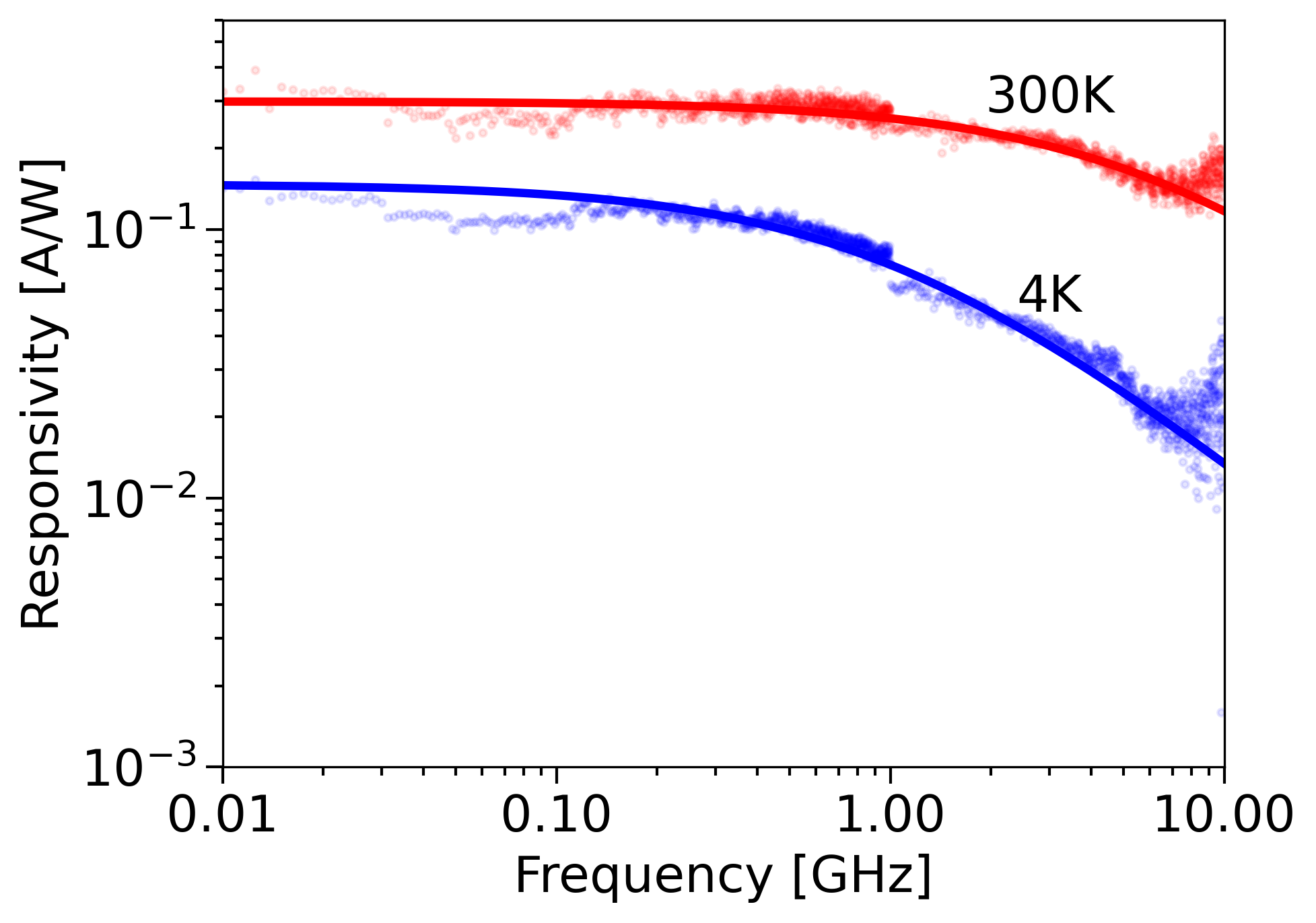}};
      \node[inner sep=0pt] (d) at (-25pt,-25pt) {\includegraphics[width=0.25\linewidth]{Figures/float_cartoon.png}};
      \node[inner sep=0pt] at (-25pt, -40pt) {\it{Floating}};
      \node[inner sep=0pt] at (-92pt, 60pt) {\bf{(b)}};
    \end{tikzpicture}
  \end{minipage}
  \caption{Frequency response for optical-microwave downconversion of the (a) vertical PWB-GePD  at $\sim 1308$~nm, away from the dropout from the microring resonator, at both 4~K and 300~K. Solid red and blue points: raw data. Solid lines: best fit with a 4th order filter from 1~GHz to 7~GHz. A resonance around 6~GHz is observed, and (b) for floating GePD. Solid red and blue points: raw data. Solid lines: best fit to first-order low-pass response. The cut off frequency is 2.7~GHz at room temperature and 400~MHz at 4.2~K. Data in (b) was measured using the same amplifier as (a) for the frequency range 1-10~GHz, and using a low-frequency amplifier between 10~MHz and 1.0~GHz.}
  \label{fig:rf}
\end{figure}

The measured optical-microwave downconversion is shown in Figure \ref{fig:rf} for a wavelength of $\sim 1308$~nm. For the vertical GePD, we observe a flat response at frequencies below 4~GHz, and a resonant enhancement of photoresponse centred at around 6~GHz. The floating GePD in comparison acts as a low-pass filter with a 3~dB bandwidth of around 400~MHz, based on a least-squares fit. The low-pass behaviour of the floating diode could be due to a high series resistance, forming a low-pass filter together with the GePD junction capacitance, consistent with the sub-exponential forward-bias IV characteristics that also suggest a high series resistance (Figure~\ref{fig:iv}). At 1~GHz, the absolute microwave photoresponse for the vertical GePD is somewhat lower than the floating GePD, both at room temperature and 4.2~K, consistent with the low-frequency response. The effective responsivity of the downconversion at lower frequencies is $300\pm 26$~mA/W ($240\pm 21$~mA/W) for the floating (vertical) GePD at room temperature, and $150 \pm 13$~mA/W ($40 \pm 3$~mA/W) at 4.2~K (Table~\ref{tab:parameters_table}). These values are very similar to the low-frequency responsivity measured without the EOM, as expected. The peaked response of the vertical diode is $410 \pm 35$~mA/W at room temperature and $150 \pm 13$~mA/W at 4.2~K.

Measured parameters of two different GePD assemblies and their respective I-V, optical and microwave characteristics are summarized in Table \ref{tab:parameters_table}. The low-frequency and microwave responses are roughly within a factor of two of each other. We attribute the small disagreement to uncertainty in residual optical mode-beating in the waveguide not corrected for by adjusting the polarization controller. This problem could be mitigated by designing future chips with single-mode waveguides at the $\sim$1310~nm range.

\begin{table*}[ht]
\centering
\caption{Parameters of the two GePD assemblies found from experiment. $I_{\text{dark}}$ is the dark current at -0.5~V, $R_{\text{DC}}$ is the peak DC responsivity over the measured wavelength sweep range. $R_{\text{MW}}$ is the microwave frequency response of photocurrent, and is reported at low frequency (within the device bandwidth), at the frequency where the response is maximized. $f_{3dB}$ is the 3~dB bandwidth.}
\begin{tabular}{|l|c|c|c|c|} 
\hline 
\textbf{Assembly} & \multicolumn{2}{c|}{\textbf{Floating}} & \multicolumn{2}{c|}{\textbf{Vertical}} \\ 
\hline 
\textbf{Temperature} & 300~K & 4.2~K & 300~K & 4.2~K  \\ 
\hline 
$I_{\text{dark}}$ (\si{\ampere}) & $3.1 \times 10^{-9}$ & $ 5.2 \times 10^{-13}$ & $1.3 \times 10^{-8} $ & $1.4 \times 10^{-10}$ \\ 
\hline 
$R_{\text{DC}}$ (A/W) & $250 \pm 11$ & $80 \pm 4$ & $270 \pm 12$ & $40 \pm 2$ \\ 
\hline
$R_{\text{MW}}$, low-frequency, 1308~nm (A/W) & $300 \pm 26$ & $150 \pm 13$ & $240 \pm 21$ & $40 \pm 3$ \\ 
\hline
$R_{\text{MW}}$, peak, 1308~nm (A/W) & $300\pm 26$ & $150 \pm 13$ & $410 \pm 35$ & $150 \pm 13$ \\ 
\hline 
$f_{\SI{3}{\deci\bel}}$ (\si{\mega\hertz}) & $2700$ & $400$ & $>6000$ & $>6000$ \\ 
\hline 
\end{tabular}
\label{tab:parameters_table}
\end{table*}

The resonant enhancement of the frequency response was observed for two different vertical GePDs that we measured. To shed light on the frequency response, we employed a microwave probe station to measure the bandwidth of a single vertical GePD on chip at room temperature, without wirebonds. This resulted on a near flat response past 10~GHz, contrasting the electrically packaged GePDs. This suggests that the resonance and low pass filter effect is due to the packaging and wire-bonding of the assemblies in Figure~\ref{fig:rf}. Indeed, the small bond-pad of the GePD allowed only a single electrical wirebond, and the presence of the shim and lack of recess in the PCB made the wirebond longer than desired, around 3~mm, so that it would have a non-negligible inductance seen by the GePD, leading to the resonance.

\section{\label{discussion}Discussion}

Having characterized the performance our PWB-GePD subsystem, we now discuss prospects for its use in qubit control. As highlighted in the previous section, the absolute room-temperature responsivity of our floating GePD-PWB subsystem is around $250\pm 13$~mA/W, lower than anticipated by around 4x compared to a subsystem with 100~\% quantum efficiency, at these wavelengths. Furthermore, the measured low-temperature responsivity is somewhat lower than the room temperature values. We expect that the the majority of suppression of the response, compared to the ideal response, is due to PWB loss, though we cannot rule out temperature-dependent loss in the GePD itself. Importantly, the optical-microwave downconversion performance of the diodes are in good agreement with the low-frequency response, given the presence of the mode beating in the waveguide at the diode operation frequency.  The unoptimized responsivity of the PWB-GePD subsystem will increase the optical power necessary to drive qubits that respond to voltages (via capacitive couplings). For a lower responsivity, the GePD will produce a smaller photo-current and drive voltage for a given optical power, and therefore, a slower qubit manipulation rate. Higher optical powers will be required to maintain the same qubit manipulation rate, which will increase overall dissipation. 

Future work using these circuits to control capacitively driven qubits with the minimum amount of optical excitation should therefore focus on improving system responsivity (A/W) towards unity quantum efficiency. This could be pursued by optimizing the entire assembly and the shape of the PWB. In future systems including loopback structures would allow for the ability to isolate the effect of the PWB and other integrated photonics devices more easily from the rest of the system to help attain lower losses and reach the performance reported in Reference \onlinecite{Lin2023}. Without this, it is very difficult to isolate the response of the photodiode from the response of the integrated photonic devices and interconnect.  

Further investigation of the impact of diode structure on performance may also help improve efficiency. It was noted in Section \ref{iv_characteristics} that the N doping density of the floating GePD is lower than that of the vertical GePD. Increasing the Si doping to N++ at the Si N+/intrinsic Ge heterojunction of the floating diode may decrease series resistance and improve bandwidth at lower temperatures. However, increasing the doping in the N region may degrade response by increasing optical loss in optical waveguide which the optical signal must pass through\cite{Fard2016}. Another investigation could be done on the effect of length on the response of the photodiode, as the increase in length would promote the absorption of light with a trade off its in RF response with the increased capacitance. 

Strategies to reduce optical excitation, while maintaining fast qubit drive, will be of interest. It has been suggested that increasing the impedance of the control line can increase the voltage drive per unit incident optical power and per unit generated electrical current by the diode\cite{Lecocq2021}. Notably, increasing the impedance will also increase the coupling and energy loss rate of the qubit to the control line, since the quantum mechanical zero point voltage fluctuations increase with the square root of the waveguide impedance. Increasing impedance of the control line therefore has the same impact on qubit loss as increasing the capacitive coupling to the qubit, while holding the control line impedance constant.  A possible means of avoiding the extra loss while maintaining higher coupling has been demonstrated in recent experiments that employ a second transmon as a saturable filter on the control waveguide. This allows classical signals to pass from the waveguide to qubit but suppresses qubit relaxation to the control line \cite{kono2020breaking}. A similar strategy employed here could help maintain qubit lifetime while reducing the optical power needed to manipulate qubits, by allowing increases in control line impedance or coupling capacitance. These aspects are related to system architecture, an important but separate topic of study, which is already receiving some attention\cite{Joshi2022}.

While we have demonstrated continuous wave optical-to-microwave conversion, almost all applications will operate with pulsed conversion. We briefly discuss pulsed operation in the context of a circuit model employing a current source, a nonlinear diode resistor, the junction capacitance, and a waveguide/qubit load. Assuming that a pulse with average power $P$ on a regular $Z=50$~$\Omega$ waveguide is sufficient for qubit drive, we arrive at a power consumption of $P_p=(\eta/Z_L R)\sqrt{2PZ}$ associated with the pulse. Here, $Z_L$ is the characteristic impedance seen by the diode, $\eta$ is the duty cycle of the control pulses, and $R$ is the diode responsivity. For concreteness, we assume $P=-80$~dBm launched on a 50-ohm waveguide suffices\cite{Lecocq2021}, an optimized responsivity $R=1$~A/W, and that the light incident on the diode is turned off between pulses, which requires a more complex setup for the optical modulator, compared to Fig.~\ref{fig:experimental_setup} and proof-of-principle demonstrations\cite{Lecocq2021}. The power dissipated in the diode from pulses alone is $P_p=60$~nW for $Z_L=50$~$\Omega$ and $P_p=0.6$~nW for $Z_L=5000$~$\Omega$ for ten percent duty cycle. These estimates are in line with other work\cite{Lecocq2021, Joshi2022} showing a thermal budget that would permit setups with thousands of converters at mK temperatures and millions of converters at Kelvin temperatures.

\section{\label{conclusion}Conclusion}

We investigated the performance of two different types of silicon-integrated germanium PIN photodiodes fabricated by a commercial foundry, connected to an optical fiber array using a polymer PWB.  The dark I-V characteristics, optical responsivity, and microwave response were measured at temperatures of 4.2~K  and 300~K. The dark I-V characteristics of both GePDs demonstrate rectifying behaviour, and the reverse bias dark current decreases substantially at 4.2~K. The vertical GePD shows a dark current of 13~nA at 300~K which decreases to 0.14~nA at 4.2~K, while the floating GePD shows a dark current of 3.1~nA at 300~K which  decreases to 0.5~pA at 4.2~K. The maximum net responsivity (that includes optical losses in the PWB) was found to be $\sim 250$~mA/W (270~mA/W) at 300~K for the floating (vertical) photodiode types. At room temperature the optical response spectrum of the floating GePD is similar in the O and C bands, but the vertical GePD has a stronger response in the C band.  The C band responsivity of both GePD types drops dramatically at 4.2~K due to the shifting bandgap of Ge.  At 4.2~K, the maximum net responsivity for microwave-to-optical conversion reduces to approximately $150 \pm 13$~mA/W and $150\pm 13$mA/W for the two different GePD types. The floating GePD exhibits evidence of a high series resistance at low temperature and a lower bandwidth 400~MHz compared to the vertical GePD which has a bandwidth exceeding 6 GHz. Better responsivity and bandwidth can be expected by refining the integrated circuit and the packaging designs informed by this study. Specifically, the photonic circuit elements can be optimized for operation in the O band, the PWB assembly can be thermally engineered to minimize thermal stress, and the germanium GePD geometry, doping profile, and the electrical contact geometry can be optimized for better high frequency performance. Future work will investigate control of electrically driven qubits using the GePDs, connected via appropriate electrical circuits. 

\begin{acknowledgments}
This work was undertaken with support from the Stewart Blusson Quantum Matter Institute (SBQMI) and the Canada First Research Excellence Fund, Quantum Materials and Future Technologies Program. We acknowledge financial support from MITACs Accelerate (IT32005), Syniad Innovations, the Canada Foundation for Innovation (CFI) and British Columbia Knowledge Development Fund (BCKDF) (Pr. 36018, 38253 and 44213) and National Science and Engineering Research Council of Canada (NSERC) (RGPIN-2019-04150, ALLRP 578461-22, CREATE 543245-2020). We acknowledge CMC Microsystems for the provision of products and services that facilitated this research, including fabrication services using the silicon photonics technology from AMF. The authors acknowledge J. Mutus and M. Hamood for valuable discussions. 
\end{acknowledgments}

\section*{Author Declarations}
\subsection*{Conflict of Interest}
S.S. and L.C. are founders of Dream Photonics.
\subsection*{Author Contributions}
\noindent
\textbf{Daniel Julien-Neitzert:} Data curation (equal); Formal analysis (equal); Investigation (lead); Methodology (equal); Software (equal); Validation (lead); Visualization (supporting); Writing - original draft (equal); Writing - review \& editing (supporting). \textbf{Edward Leung:} Data curation (equal); Formal analysis (equal); Investigation (supporting); Methodology (equal); Software (equal); Validation (supporting); Visualization (lead); Writing - original draft (equal); Writing - review \& editing (supporting). \textbf{Nusair Islam:} Formal analysis (supporting); Investigation (supporting); Methodology (supporting); Writing - review \& editing (supporting). \textbf{Serge Khorev:} Formal analysis (supporting); Methodology (supporting); Project Administration (lead); Supervision (supporting); Writing - original draft (supporting); Writing - review \& editing (supporting). \textbf{Sudip Shekhar:} Conceptualization (equal); Formal analysis (supporting); Funding acquisition (equal); Methodology (supporting); Resources (supporting); Supervision (equal); Writing - original draft (supporting); Writing - review \& editing (supporting). \textbf{Lukas Chrostowski:} Conceptualization (equal); Formal analysis (supporting); Funding acquisition (equal); Methodology (supporting); Resources (equal); Supervision (equal); Writing - original draft (supporting); Writing - review \& editing (supporting). \textbf{Jeff F. Young:} Conceptualization (supporting); Formal analysis (supporting); Funding acquisition (equal); Methodology (supporting); Resources (supporting); Supervision (supporting); Writing - original draft (supporting); Writing - review \& editing (supporting). \textbf{Joe Salfi:} Conceptualization (equal); Formal analysis (supporting); Funding acquisition (equal); Methodology (supporting); Resources (equal); Supervision (equal); Writing - original draft (supporting); Writing - review \& editing (lead). 
 
\section*{Data Availability Statement}
The data that support the findings of this study are available from the corresponding author upon reasonable request. 

\section*{References}
\nocite{*}
%

\appendix
\section{System Losses and Calibration}
\begin{figure}[ht!]
	\centering
	\includegraphics[width=7cm]{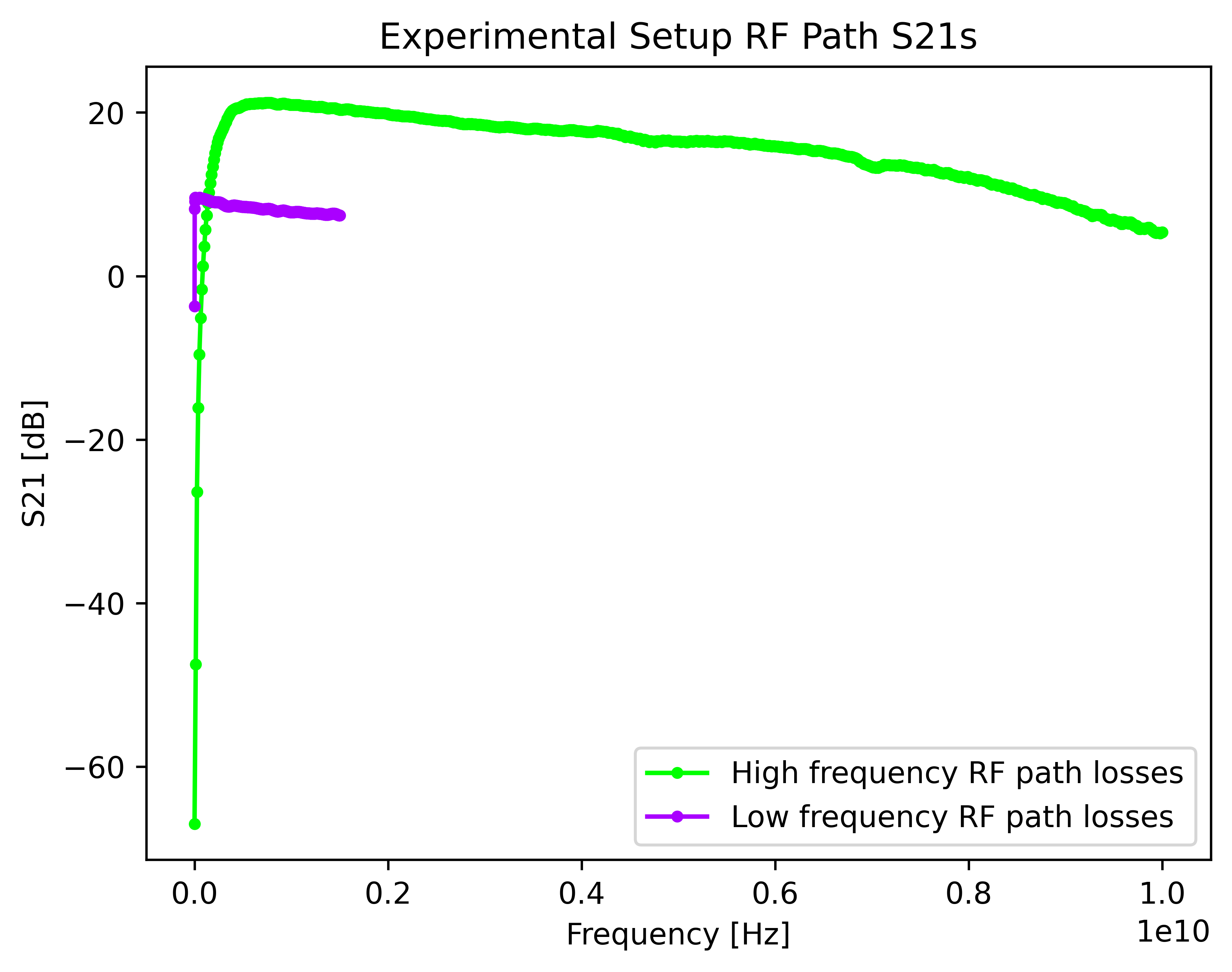}
	\caption{Plot of the frequency dependant losses and gains in the microwave measurement system summed together. This captures the losses of most microwave frequency dependant components including the microwave cables. The gain of the amplifier was also captured as it has as strong frequency dependence. On chip and PCB electrical connections such as wirebonds and transmission lines were not measured}
 \label{rf_path}
\end{figure}

\begin{table}
    \centering
    \caption{Estimated optical losses between the reference plane (optical fiber feedthrough) and the GePD, for both GePDs. The optical waveguide loss is 2~dB/cm. The total uncertainty loss is estimated at 1~dB from optical fiber connectors.} 
    \begin{tabular}{|c|c|c|c|}
         \hline
         Subsystem & Where & Loss (dB) & Loss (dB) \\
         & & Vertical & Float \\
         \hline\hline
         Optical fiber, connectors & Dipstick & $1.25\pm 0.2\footnotemark[1]$ & $1.25\pm 0.2\footnotemark[1]$ \\
         \hline
         Waveguide (2 dB/cm) \footnotemark[3] & On-chip & $1.23\pm 0.1\footnotemark[2]$ & $0.22\pm 0.1\footnotemark[2]$\\
         \hline
         Waveguide Y-branch\cite{Zhang:13} & On-chip  & $3.3\pm 0.02$ (O-band) & - \\
         & & $3.6\pm 0.02$ (C-band) & - \\
         \hline
         \hline
         Total & O-band & $5.78\pm 0.2$ & $1.47\pm 0.2$ \\ 
               & C-band & $6.08\pm 0.2$ & $1.47\pm 0.2$ \\
         \hline      
    \end{tabular}
    \footnotetext[1]{Estimate of uncertainty in loss from 3 fiber connectors.}
    \footnotetext[2]{Estimate of uncertainty from direct measurement of waveguide loss at room temperature.}
    \footnotetext[3]{Based on foundry data}
    \label{tab:loss_optical}
\end{table}

 \begin{table}
    \centering
    \caption{Estimated microwave losses on the input to the EOM and output to the VNA and the GePD, for both GePDs.}
    \begin{tabular}{|c|c|c|}
         \hline
         Subsystem & Where & Loss (dB, 5~GHz) \\
         \hline\hline
         CuNi semi-rigid coax (output) & Dipstick & $0.38$ \\
         \hline
         Flex coax (output) & Ambient & $4.71\pm0.1\footnotemark[1]$ \\
         \hline
         Bias Tee (output) & Ambient & $0.84$ \\
         \hline
         Amplifier (input) & Ambient  & $-20.75\pm 0.5$ \\
         \hline
         Flex coax (input) & Ambient & $0.66\pm0.1\footnotemark[1]$ \\
         \hline
         Flex coax (input) & Ambient & $0.66\pm0.1\footnotemark[1]$ \\
         \hline
         EOM calibration uncertainty (input) & Ambient & $0 \pm 0.2$\footnotemark[2]\\
         \hline
         Total & & $-14.16 \pm 0.6$\\
         \hline
    \end{tabular}
    \footnotetext[1]{Based on best estimate.}
    \footnotetext[2]{Estimated based on observed drift of EOM, but does not include EOM conversion loss, which is calibrated as described in the main text. }
    \label{tab:loss_microwave}
\end{table}

\begin{figure}[ht!]
    \centering
    \begin{tikzpicture}
      \node[inner sep=0pt] at (0,0) {\includegraphics[width=0.9\linewidth]{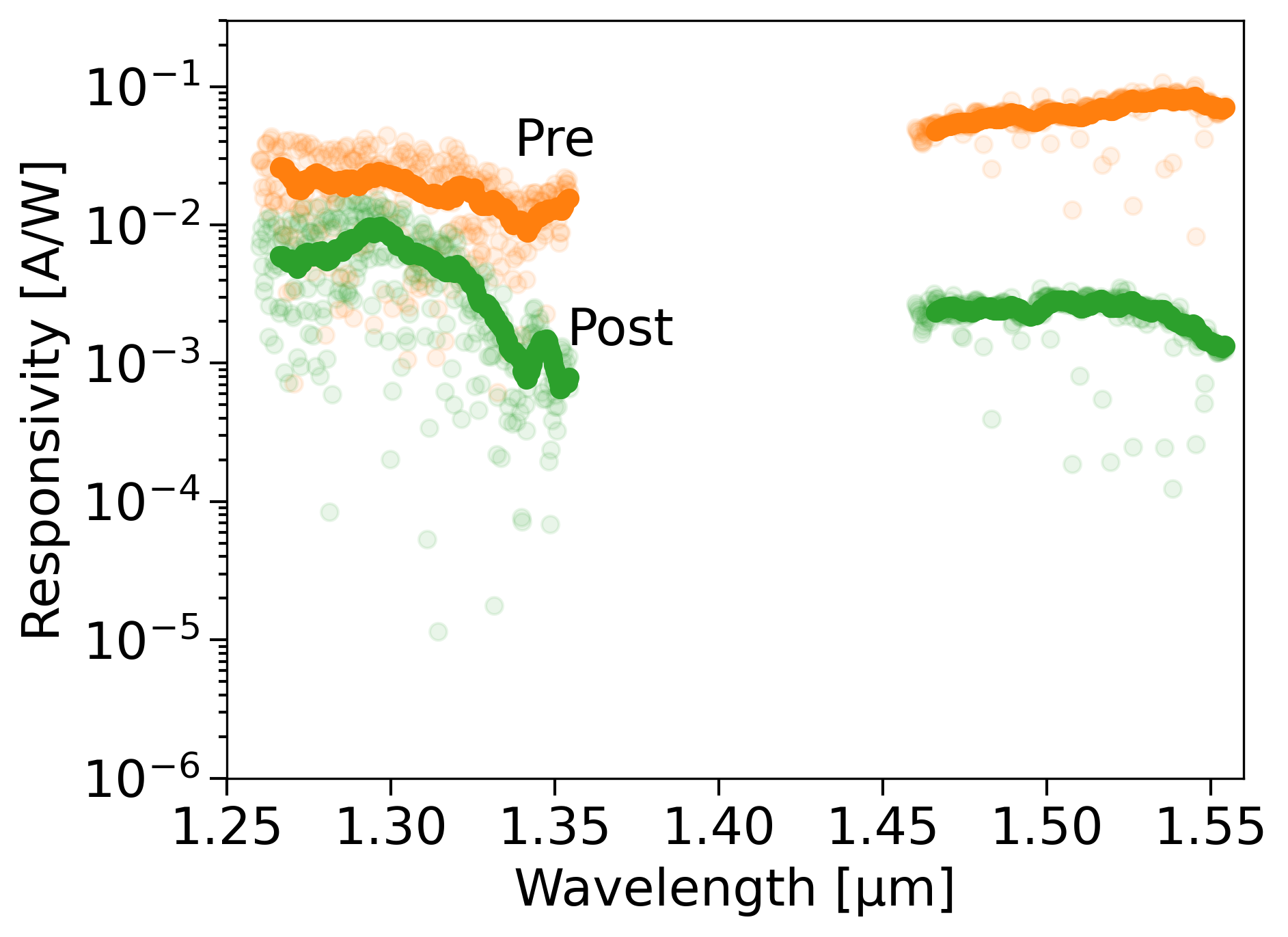}};
      \node[inner sep=0pt] at (70pt,-25pt) {\includegraphics[width=0.20\linewidth]{Figures/rect_cartoon.png}};
      \node[inner sep=0pt] at (70pt, -40pt) {\it{Vertical}};
    \end{tikzpicture}
  \caption{The optical response of the vertical GePD pre- and post- 4.2~K cooldown. There is a factor of three between the two measurement, where responsivity is at its highest in the O-band.}
  \label{fig:pre_post_ratio}
\end{figure}

Fig. \ref{fig:pre_post_ratio} plots the optical response of the vertical diode pre- and post- cooldown to 4~K. The post optical measurements were made once the assembly had completely returned to room temperature (300~K). There is a difference between the pre- and post- optical response, namely a change in the responsitivity from our photodiode. In the O-band range, where the highest responsivity was reported at 4.2~K there is a difference of a factor of three in its response at room temperature compared to its response pre-cooldown. In both C- and O- bands there seems to be an increase in the ratio between the two measurements at the longer wavelengths of each band. Interestingly, the shape of the post-cooldown optical response follows a similar shape to that of the optical response measured at 4.2~K in the O-band. This is indicative of a change in the optical path at 4.2~K as the shape of the PWB changes in such a way that its response during the cooldown follows a similar shape of the response at room temperature post-cooldown.

\section{Device Manufacturing and Assembly Details}
\label{sec:manufacturing}

\begin{figure}
 \begin{minipage}[b]{0.45\textwidth}
    \centering
    \begin{tikzpicture}
      \node[inner sep=0pt] at (0,0) {\includegraphics[width=\linewidth]{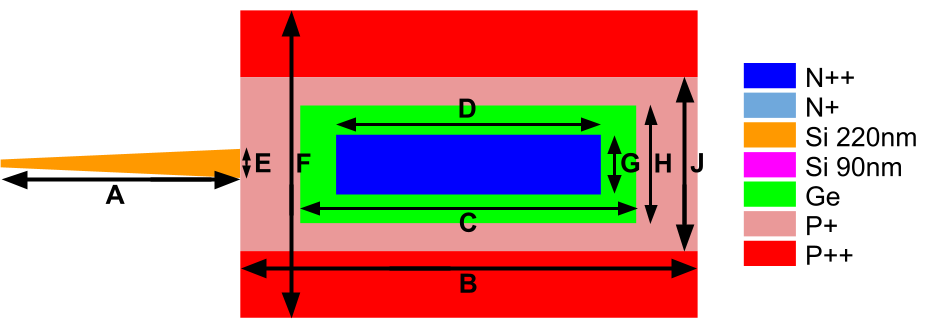}};  
      \node[inner sep=0pt] at (-110pt, 30pt) {\bf{(a)}};
    \end{tikzpicture}
  \end{minipage}
  \begin{minipage}[b]{0.45\textwidth}
    \centering
    \begin{tikzpicture}
      \node[inner sep=0pt] at (0,0) {\includegraphics[width=\linewidth]{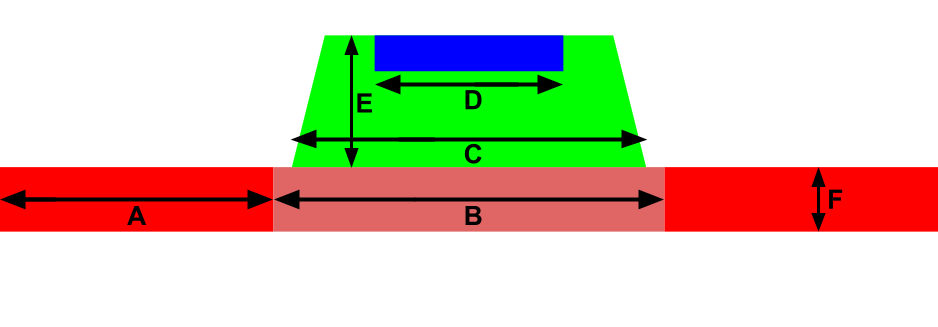}};  
      \node[inner sep=0pt] at (-110pt, 30pt) {\bf{(b)}};
    \end{tikzpicture}
  \end{minipage}
  \hfill
  \caption{Design dimensions of the vertical diode used in experiments, neglecting the metal contacts. Note that the N++ doping on top of the germanium symbolizes doped germanium whereas all other doped regions are doped silicon. Surrounding the diode on every side is silicon dioxide. Note that the drawings are not to scale. a) The top view of the diode. The silicon waveguide tapers up from a \SI{0.5}{\micro\meter} waveguide to deliver the optical signal to the diode. In this drawing A=\SI{111}{\micro\meter}, B=\SI{15}{\micro\meter}, C=\SI{11}{\micro\meter}, D=\SI{8.5}{\micro\meter}, E=\SI{2.2}{\micro\meter}, F=\SI{21}{\micro\meter}, G=\SI{4}{\micro\meter}, H=\SI{8}{\micro\meter}, J=\SI{12}{\micro\meter}. Color key is shown on the right. b) The cross section view of the diode taken at the center of the doped regions. In this drawing A=\SI{4.6}{\micro\meter}, B=\SI{12}{\micro\meter}, C=\SI{8}{\micro\meter}, D=\SI{4}{\micro\meter}, E=\SI{0.5}{\micro\meter}, F=\SI{0.22}{\micro\meter}.}   
  \label{fig:vertical_detail}
\end{figure}

\begin{figure}
  \begin{minipage}[b]{0.45\textwidth}
    \centering
    \begin{tikzpicture}
      \node[inner sep=0pt] at (0,0) {\includegraphics[width=\linewidth]{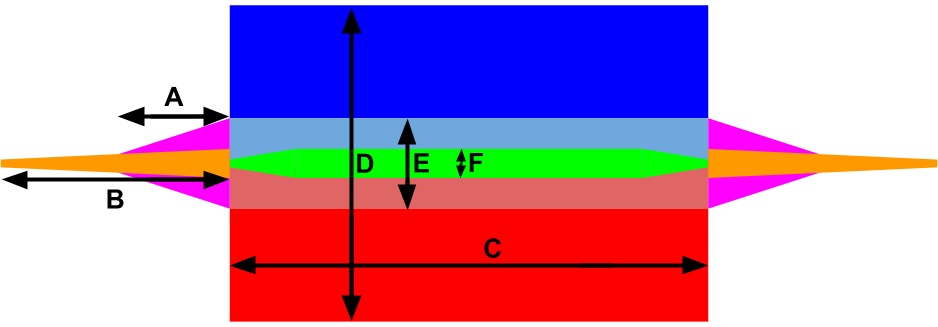}};   
      \node[inner sep=0pt] at (-110pt, 30pt) {\bf{(a)}};
    \end{tikzpicture}
  \end{minipage}
  \hfill
  \begin{minipage}[b]{0.45\textwidth}
    \centering
    \begin{tikzpicture}
      \node[inner sep=0pt] at (0,0) {\includegraphics[width=\linewidth]{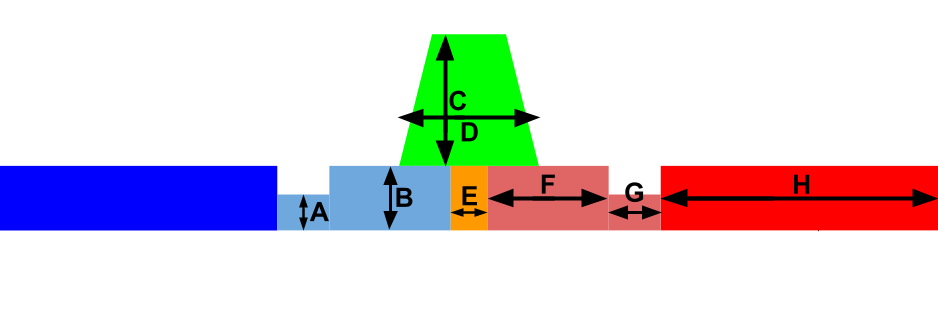}};  
      \node[inner sep=0pt] at (-110pt, 30pt) {\bf{(b)}};
    \end{tikzpicture}
  \end{minipage}
  \caption{Design dimensions of the floating diode used in experiments, neglecting the metal contacts. Color key is shown in Figure \ref{fig:vertical_detail}a). Surrounding the diode on every side is silicon dioxide. Note that the drawings are not to scale. a) The top view of the diode. The silicon waveguide tapers up from a \SI{0.5}{\micro\meter} waveguide to deliver the optical signal to the diode. Note that in this diode there is a second taper where light may exit the diode, which is attached to an integrated photonic loop mirror to allow for the light to pass through the diode twice and be more fully absorbed. In this drawing A=\SI{22}{\micro\meter}, B=\SI{50}{\micro\meter}, C=\SI{21}{\micro\meter}, D=\SI{14.25}{\micro\meter}, E=\SI{4.25}{\micro\meter}, F=\SI{1.25}{\micro\meter}. b) The cross section view of the diode taken at the center of the doped regions. In this drawing A=\SI{0.09}{\micro\meter}, B=\SI{0.22}{\micro\meter}, C=\SI{0.5}{\micro\meter}, D=\SI{1.25}{\micro\meter}, E=\SI{0.55}{\micro\meter}, F=\SI{1.85}{\micro\meter}, G=\SI{1.125}{\micro\meter}, H=\SI{5}{\micro\meter}. }
  \label{fig:floating_detail} 
\end{figure}

The Advanced Micro Foundry (AMF) integrated photonic process includes germanium and germanium dopants which allow for fabrication of integrated germanium-based photodiodes on a silicon-on-insulator platform for illumination using silicon waveguides. Design dimensions of the two diodes used are shown in Figures \ref{fig:vertical_detail} and \ref{fig:floating_detail}. Note that some dimensions are not shown because they are unknown to the designer or not available to public knowledge such as the depth of the N++ germanium implant. Both GePDs are based on designs described in ref\cite{Fard2016, Novack2013,Zhang2014}.

\begin{figure}
  \begin{minipage}[b]{0.48\textwidth}
    \begin{tikzpicture}
      \node[inner sep=0pt] at (0,0) {\includegraphics[width=\linewidth]{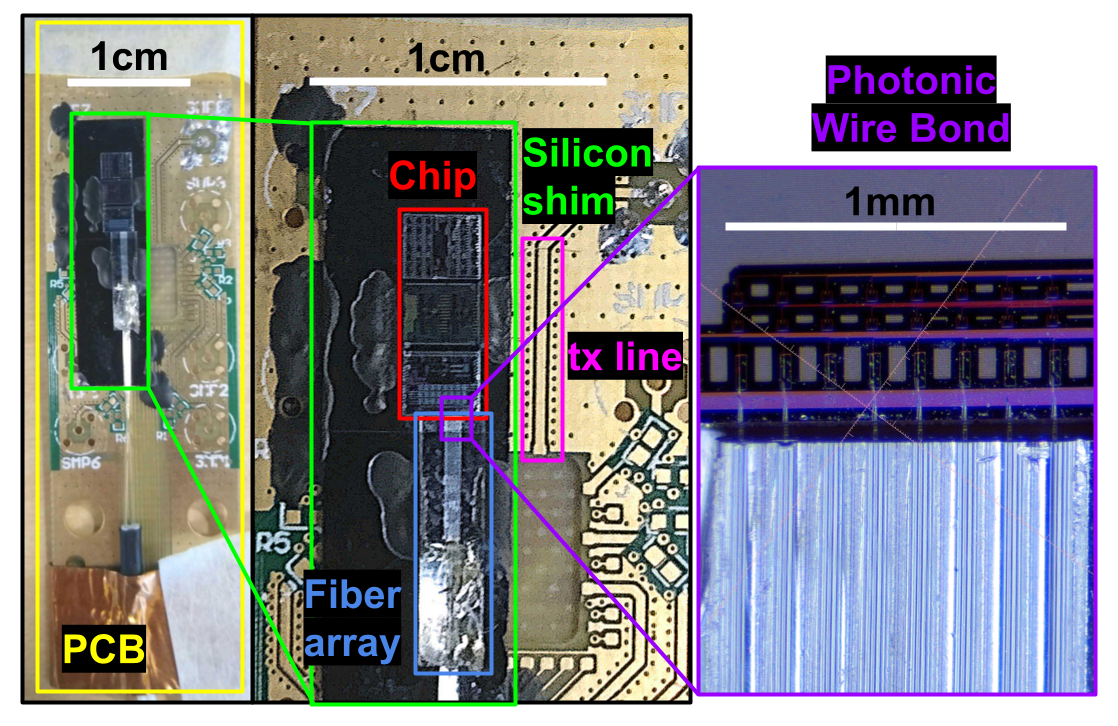}};  
      \node[inner sep=0pt] at (-110pt, 70pt) {\bf{(a)}};
    \node[inner sep=0pt] at (-58pt, 70pt) {\bf{(b)}};
    \node[inner sep=0pt] at (37pt, 70pt) {\bf{(c)}};
     \end{tikzpicture}
  \end{minipage}
  \caption{{a) A photograph of one of the two assemblies, at a scale to show the PCB and silicon shim. The PCB is approximately \SI{7}{\centi\meter} by \SI{2}{\centi\meter}. b) A closer photograph of the assembly showing the fiber array, chip, and silicon shim all glued in place. PWBs and electrical wirebonds too small to be seen at this scale attach the fiber array to the chip, and the chip to the adjacent transmission line (tx line) respectively. The chip is approximately \SI{7.5}{\milli\meter} by \SI{3}{\milli\meter}. c) A picture taken under an optical microscope showing the PWBs bridging the fiber array and the chip. Note that while multiple bonds are present, only one of them connects to the waveguide which delivers light to the tested diode. The PWBs are spaced apart by \SI{127}{\micro\meter}.}}
  \label{fig:assembly_irl} 
\end{figure}
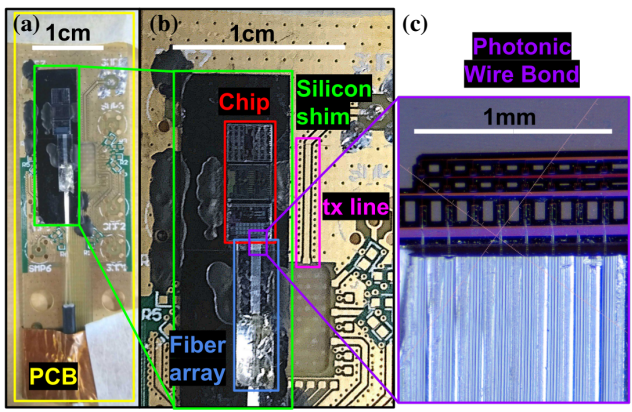

\end{document}